\documentclass{ptptex}
\usepackage{graphicx}
\usepackage{revsymb}
\usepackage{bm}

\newcommand{\sA}{\mathop{\text{s}}\nolimits_\text{A}}
\newcommand{\sB}{\mathop{\text{s}}\nolimits_\text{B}}
\newcommand{\cA}{\mathop{\text{c}}\nolimits_\text{A}}
\newcommand{\cB}{\mathop{\text{c}}\nolimits_\text{B}}
\newcommand{\ket}[1]{|{#1}\rangle}
\newcommand{\bra}[1]{\langle{#1}|}
\newcommand{\bras}[2]{{}_{#2}\hspace*{-0.2mm}\langle{#1}|}
\newcommand{\ketbras}[3]{\ket{#1}_{#3}\hspace*{-0.2mm}\bra{#2}}
\newcommand{\Tr}{\mathop{\text{Tr}}\nolimits}

\title{%        %You can use \\ for explicit line-break
A Simple Scheme to Entangle Distant Qubits from a Mixed State via an Entanglement Mediator%
}

\author{%       %Use \scshape  for the family name
Kazuya \textsc{Yuasa}\footnote{E-mail: yuasa@hep.phys.waseda.ac.jp}
and Hiromichi \textsc{Nakazato}\footnote{E-mail: hiromici@waseda.jp}%
}

\inst{%         %Affiliation, neglected when [addenda] or [errata]
Department of Physics, Waseda University, Tokyo 169-8555, Japan
}

\recdate{November 29, 2004}%            %Editorial Office will fill in this.

\abst{%         %this abstract is neglected when [addenda] or [errata]
A simple scheme to prepare an entanglement between two separated
qubits from a given mixed state is proposed. A single qubit
(entanglement mediator) is repeatedly made to interact locally and
consecutively with the two qubits through rotating-wave couplings
and is then measured. It is shown that we need to repeat this kind
of process only three times to establish a maximally entangled state
directly from an arbitrary initial mixed state, with no need to
prepare the state of the qubits in advance or to rearrange the setup
step by step. Furthermore, the maximum yield realizable with this
scheme is compatible with the maximum entanglement, provided that
the coupling constants are properly tuned. }

\begin{document}

\maketitle

Entanglement is one of the peculiar concepts in quantum physics. It
shows a nonlocal property of quantum mechanics, and is the essence 
of nonclassical phenomena. This concept is now regarded as a key to
moving beyond classical ideas and plays a central role in the field
of quantum information and computation
\cite{ref:QuantInfoCompChuang,ref:QuantInfoCompZeilinger}. The
peculiarity of entanglement reveals itself especially when it
involves spatially separated systems. Nonlocality has been studied
in such situations \cite{ref:Peres}, and such an entanglement is of
fundamental importance in quantum communication, for example,
quantum teleportation
\cite{ref:QuantInfoCompChuang,ref:QuantInfoCompZeilinger}.

Practically speaking, an important problem is how to prepare an
entanglement between distant parties, and it has been studied for
decades
\cite{ref:Haroche1997,ref:Bergou1997,ref:CiracZollerEntGeneDistant,ref:EntanglingDistantAtoms2003,ref:ZeilingerTeleportationDanube}.
In particular, it is often required to prepare such an entanglement
from a (naturally) given state, which is in general a mixed state.
Entanglement distillation
\cite{ref:QuantInfoCompZeilinger,ref:PurificationBennettPRL1996,ref:TYamamotoNature}
is therefore an important topic in this context.

A novel mechanism for driving a quantum system from an arbitrary
mixed state to a pure state, which makes use of Zeno-like
measurements,\cite{ref:qpf,ref:qpfe} was discovered recently. This
mechanism is that responsible for the following types of behavior: A
series of repeated measurements applied to a quantum system can
(asymptotically) drive a second system that interacts with the first
one into a pure state, even if the second system is initially in an
arbitrary mixed state.\footnote{Measurements play an important role
in the scheme proposed in
Refs.~\citen{ref:qpf,ref:qpfe,ref:qpfeLidar}, and in this regard,
the mechanism discussed here seems to be closely related to the
so-called quantum Zeno effect (QZE) \cite{ref:QZE}. But this
mechanism is quite different, as pointed out in
Ref.~\citen{ref:qpf}: In contrast to the fact that a small time
interval between measurements is required in the QZE, the time
intervals in the present scheme need not be small but, rather, are
treated as parameters to be adjusted in such a manner to realize an
efficient preparation of a pure state.} This provides us with useful
procedures for preparing an entanglement between qubits,
initialization of multiple qubits,
etc.~\cite{ref:qpfe,ref:qpfeLidar}. Furthermore, it has been
extended in Ref.~\citen{ref:qpfes} to a scheme to establish an
entanglement between \textit{distant} systems, where measurements
are repeated on an ``entanglement mediator.''

In this article, we point out that further improvement is possible
for the scheme proposed in Ref.~\citen{ref:qpfes} in the case that a
single qubit (entanglement mediator) interacts with each of the two
(spatially separated) qubits, between which an entanglement is
prepared, through a rotating-wave coupling, which is often discussed
in the literature
\cite{ref:Haroche1997,ref:Bergou1997,ref:CiracZollerEntGeneDistant,ref:EntanglingDistantAtoms2003}.
We show in the following that it is indeed possible to obtain a
maximally entangled state from an arbitrary mixed state
\textit{without repeating measurements many times}, if the scheme
presented in Ref.~\citen{ref:qpfes} is modified properly. In fact,
it is found that we actually need only repeat the process (i.e., the
interactions of the mediator with the qubits and the measurement of
the mediator) three times. Furthermore, the maximum yield attainable
in this scheme is shown to be compatible with the maximum
entanglement.

\begin{figure}[t]
\centerline{\includegraphics[width=0.5\textwidth]{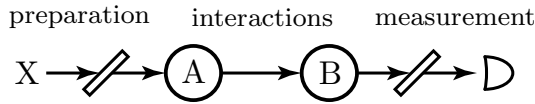}}
\caption{A scheme for entanglement generation via successive
interactions of an entanglement mediator X with qubits A and B\@.}
\label{fig:SeqInt}
\end{figure}

We consider two qubits, A and B, which are spatially separated and
do not directly interact with each other. In order to make them
entangled, another two-level system (the entanglement mediator), X, 
prepared in a particular state is caused to interact with A and B
successively and is measured (Fig.~\ref{fig:SeqInt}).
Such entangling schemes are studied in Ref.~\citen{ref:Bergou1997},
but there it is assumed that the initial states of the distant
qubits A and B are prepared in specific pure states. What we show in
the following is that no preparation is necessary for A and B\@.
Indeed, a maximally entangled state can be obtained directly from a
given (mixed) state \textit{within the same (fixed) setup}, without
introducing other resources or special tunings/arrangements for the
initialization of the qubits.

It should be stressed here that the present scheme assumes the
application of nonlocal operations of the qubits A and B indirectly
through the operations applied to the mediator X, and hence, it is
beyond the realm of the entanglement distillation discussed in
Refs.~\citen{ref:QuantInfoCompZeilinger,ref:PurificationBennettPRL1996,ref:TYamamotoNature},
etc., which allow only local operations and classical communication.

We assume that the qubits A and B and the entanglement mediator X
are two-level systems with a common energy gap $\Omega$ and that
each qubit interacts separately with the mediator through a
rotating-wave coupling. The free evolution of the three systems is
hence described by the Hamiltonian\footnote{In this article, we
consider the idealized case in which the effect of dissipation is
negligible.}
\begin{equation}
H_0=\frac{\Omega}{2}(
\sigma_3^\text{(A)}
+\sigma_3^\text{(B)}
+\sigma_3^\text{(X)}),
\end{equation}
and while A (B) interacts with X for a time interval
$\tau_\text{A(B)}$, the total system evolves according to the
Hamiltonian
\begin{equation}
H_0+H_\text{XA(B)},
\label{eqn:DuringInteraction}
\end{equation}
with
\begin{subequations}
\label{eqn:Ints}
\begin{gather}
H_\text{XA}
=g_\text{A}
(\sigma_+^\text{(X)}\sigma_-^\text{(A)}
+\sigma_-^\text{(X)}\sigma_+^\text{(A)}), \label{eqn:IntXA}\\
H_\text{XB}
=g_\text{B} (\sigma_+^\text{(X)}\sigma_-^\text{(B)}
+\sigma_-^\text{(X)}\sigma_+^\text{(B)}). \label{eqn:IntXB}
\end{gather}
\end{subequations}
Here, $\sigma_i$ ($i=1,2,3$) are the Pauli operators, and
$\sigma_\pm=(\sigma_1\pm i\sigma_2)/2$ are the ladder operators. The
coupling constant $g_\text{A(B)}$ between the qubit A (B) and
mediator X is assumed to be an adjustable parameter, \textit{but it
is fixed throughout the process} described in the following.

It is actually possible to initialize A and B in advance, as
required in Ref.~\citen{ref:Bergou1997}, without introducing
additional resources in the above setup: One can transfer the pure
state $\ket{\uparrow}$ of X (which is the eigenstate of the operator
$\sigma_3$ with eigenvalue $+1$) to A (B) by applying the
Hamiltonian (\ref{eqn:DuringInteraction}) for an appropriate time
interval $g_\text{A(B)}\tau_\text{A(B)}=\pi/2$, and the qubits are
thereby prepared in the pure state
$\ket{\uparrow\uparrow}_\text{AB}$. Note, however, that special
arrangements of the values of $g_\text{A}\tau_\text{A}$ and
$g_\text{B}\tau_\text{B}$, which differ from those necessary for
entangling qubits maximally, are required solely for this
initialization. It is shown that, without rearranging these
parameter values but, rather, fixing them throughout the entire
process, we can prepare a maximally entangled state directly from a
mixed state in the present scheme.

The idea for obtaining an entanglement from an arbitrary mixed state
$\varrho_\text{AB}$ is the following. An important point is that the
excitation number of qubits
\begin{equation}
N=\frac{1+\sigma_3^\text{(A)}}{2}
+\frac{1+\sigma_3^\text{(B)}}{2}
+\frac{1+\sigma_3^\text{(X)}}{2}
\label{eqn:ExcitationNo}
\end{equation}
is a constant of motion, due to the rotating-wave couplings in
(\ref{eqn:Ints}). We first prepare the mediator X in the state
$\ket{\uparrow}_\text{X}$\footnote{Such a preparation is not very
difficult for a particular choice of X\@. For example, if X is a
photon, as discussed below, one can easily prepare its polarization
using a polarizer.} and let it interact with the qubits A and B
successively, through the interactions (\ref{eqn:IntXA}) and
(\ref{eqn:IntXB}) for the time intervals $\tau_\text{A}$ and
$\tau_\text{B}$, respectively. After these consecutive interactions,
the state of X is measured to see whether it has flipped down into
the state $\ket{\downarrow}_\text{X}$. If it has, we know that one
and only one of the qubits has certainly flipped from
$\ket{\downarrow}$ to $\ket{\uparrow}$, since the excitation number
$N$ given in (\ref{eqn:ExcitationNo}) is conserved in this process.
This process, that is, the preparation of X in the state
$\ket{\uparrow}_\text{X}$, the interactions between X and the two
qubits A and B, and the projection to $\ket{\downarrow}_\text{X}$,
entails the following state changes of the initial state
$\varrho_\text {AB}$:
\begin{subequations}
\begin{equation}
\begin{array}{ccc}
\medskip
\ket{\uparrow\uparrow}_\text{AB}&\longrightarrow&{\rm disappearing},\\
\medskip
\ket{\uparrow\downarrow}_\text{AB}\quad\text{and}\quad
\ket{\downarrow\uparrow}_\text{AB}
&\longrightarrow&\ket{\uparrow\uparrow}_\text{AB},\\
\ket{\downarrow\downarrow}_\text{AB}
&\longrightarrow&\ket{\uparrow\downarrow}_\text{AB}\quad\text{or}\quad
\ket{\downarrow\uparrow}_\text{AB}.
\end{array}
\end{equation}
Note that the $\ket{\uparrow\uparrow}_\text{AB}$ component of the
initial state $\varrho_\text {AB}$ is eliminated and the
$\ket{\downarrow\downarrow}_\text{AB}$ sector becomes vacant after
the measurement (projection). Next, a second mediator X prepared in
the state $\ket{\uparrow}_\text{X}$ is caused to interact with A and
B (not necessarily in the same order as the first process), and it
is subsequently confirmed again to be in the state
$\ket{\downarrow}_\text{X}$. In this second process too, it is
possible that the state of only one of the qubits has flipped from
$\ket{\downarrow}$ to $\ket{\uparrow}$, and the
$\ket{\uparrow\uparrow}_\text{AB}$ component has thus been
eliminated:
\begin{equation}
\begin{array}{ccc}
\medskip
\ket{\uparrow\uparrow}_\text{AB}&\longrightarrow&{\rm disappearing},\\
\ket{\uparrow\downarrow}_\text{AB}\quad\text{and}\quad
\ket{\downarrow\uparrow}_\text{AB}
&\longrightarrow&\ket{\uparrow\uparrow}_\text{AB}.
\end{array}
\end{equation}
Now, only the $\ket{\uparrow\uparrow}_\text{AB}$ sector is occupied.
This was originally the $\ket{\downarrow\downarrow}_\text{AB}$
component of the initial state $\varrho_\text{AB}$. That is, the
state $\ket{\downarrow\downarrow}_\text{AB}$ has been extracted from
the initial state $\varrho_\text{AB}$ in the form of
$\ket{\uparrow\uparrow}_\text{AB}$, and the qubits A and B have thus
been driven to the pure state $\ket{\uparrow\uparrow}_\text{AB}$
through these two processes. After this ``purification,'' the third
mediator X prepared in the state $\ket{\downarrow}_\text{X}$ is
caused to interact with A and B in a prescribed order, and the state
$\ket{\uparrow}_\text{X}$ is confirmed. This, in turn, flips
one of the qubits (but we do not know which one) from
$\ket{\uparrow}$ to $\ket{\downarrow}$,
\begin{equation}
\begin{array}{ccc}
\ket{\uparrow\uparrow}_\text{AB}
&\longrightarrow&
\ket{\uparrow\downarrow}_\text{AB}\quad\text{or}\quad
\ket{\downarrow\uparrow}_\text{AB},
\end{array}
\end{equation}
\end{subequations}
and an entanglement between the distant qubits A and B results.
Thus, an entanglement can be obtained from an arbitrary mixed state
$\varrho_\text{AB}$ after only three measurements, two for the
purification and one for the creation of the entanglement.  Also,
note that this is done with a fixed parameter set, i.e., without
readjusting the interaction strengths for the creation of the
entanglement. Furthermore, there is no need to repeat these 
processes many times. Of course, we still need to optimize the 
processes in order to create an entanglement efficiently, as is 
clarified below.

There are several possible recipes for this entanglement
preparation, because we have the freedom to choose the order in
which the mediator X interacts with the qubits A and B for each
process.  Actually, there are essentially four combinations of such
orders. (We can fix the order of the interactions with the mediator
in the last process without loss of generality. Then, we are left
with two choices for the first mediator, and two choices for the
second.) The important point here is to understand which recipe
(i.e., the combination of orders of the interactions) enables us to
obtain a highly entangled state with a high yield. In the
present scheme described in the preceding paragraph, every one of
the three measurements applied to the entanglement mediators
extracts the correct outcome, and they therefore produce an
entangled state with certainty.  This is due to the existence of the
conserved quantity $N$ given in (\ref{eqn:ExcitationNo}), according
to which all states are classified into four sectors with different
values of $N$, namely $N=0,1,2$, and $3$, and to the combination of
three measurements (projections), which only one of the sectors
survives.  However, we still need to clarify which combination of
orders is the most efficient one, in the sense that the desired
results are realized with high probability, and at the same time, a
highly entangled state (i.e., a state whose concurrence
\cite{ref:Concurrence} is close to unity) is obtained.

\begin{figure}
\centerline{\includegraphics[width=0.5\textwidth]{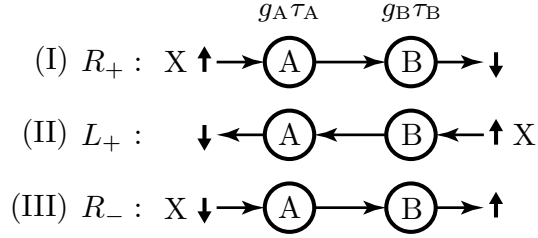}}
\caption{A recipe for optimal entanglement preparation.
The first mediator moves rightward (I), the second one moves
leftward (II), and the last one moves rightward (III)\@. (Here,
``rightward" means that the interaction between X and A is followed
by that between X and B, and ``leftward" means that the order is
opposite.) If the mediators are found to be in the states indicated
in this figure, the processes (I)--(III), represented by the
operators $R_+$, $L_+$, and $R_-$, respectively, extract a specific
(entangled) state of the qubits A and B\@. The strength of the
interaction between X and A (B) is fixed
$g_\text{A(B)}\tau_\text{A(B)}$ throughout the processes
(I)--(III).}
\label{fig:SeqInts}
\end{figure}
It is elementary to confirm that the optimal preparation of a
maximally entangled state is possible via the recipe presented in
Fig.~\ref{fig:SeqInts}.  In the situation depicted there, an
appropriate choice of the parameters $g_\text{A}\tau_\text{A}$ and
$g_\text{B}\tau_\text{B}$ enables us to obtain a maximally entangled
state (with concurrence $1$) with the maximum yield realizable in
the present scheme. The mediator X is first allowed to travel
rightward, that is, to interact with the qubit A and then B\@. If it
is found to be in the state $\ket{\downarrow}_\text{X}$ after the
interactions with A and B on the right, the state of A and B changes
according to
\begin{equation}
\varrho_\text{AB}
\to R_+\varrho_\text{AB}R_+^\dag/
\Tr_\text{AB}(R_+\varrho_\text{AB}R_+^\dag),
\label{eqn:FirstStep}
\end{equation}
where
\begin{align}
R_+&\equiv\bras{\downarrow}{\text{X}}
e^{-iH_\text{XB}\tau_\text{B}}
e^{-iH_\text{XA}\tau_\text{A}}
\ket{\uparrow}_\text{X}\nonumber\displaybreak[0]\\
&={-i}\,\Bigl[
\sA\ketbras{\uparrow}{\downarrow}{\text{A}}
\otimes\Bigl(
\ketbras{\downarrow}{\downarrow}{\text{B}}
+\cB\ketbras{\uparrow}{\uparrow}{\text{B}}
\Bigr)
+\sB\,\Bigl(
\ketbras{\uparrow}{\uparrow}{\text{A}}
+\cA\ketbras{\downarrow}{\downarrow}{\text{A}}
\Bigr)\otimes\ketbras{\uparrow}{\downarrow}{\text{B}}
\Bigr]
\end{align}
in the interaction picture. (Here, $\sA\equiv\sin
g_\text{A}\tau_\text{A}$, $\cA\equiv\cos g_\text{A}\tau_\text{A}$,
etc.)  The second and third mediators are translated leftward and
rightward, as depicted in Fig.~\ref{fig:SeqInts}, and as a result,
the state of the qubits A and B changes further.  The relevant
operators read
\begin{align}
L_+&\equiv\bras{\downarrow}{\text{X}}e^{-iH_\text{XA}\tau_\text{A}}
e^{-iH_\text{XB}\tau_\text{B}}\ket{\uparrow}_\text{X}\nonumber\\
&={-i}\,\Bigl[
\sA\ketbras{\uparrow}{\downarrow}{\text{A}}\otimes\Bigl(
\ketbras{\uparrow}{\uparrow}{\text{B}}
+\cB\ketbras{\downarrow}{\downarrow}{\text{B}}
\Bigr)
+\sB\,\Bigl(
\ketbras{\downarrow}{\downarrow}{\text{A}}
+\cA\ketbras{\uparrow}{\uparrow}{\text{A}}
\Bigr)\otimes\ketbras{\uparrow}{\downarrow}{\text{B}}
\Bigr]
\intertext{and}
R_-&\equiv\bras{\uparrow}{\text{X}} e^{-iH_\text{XB}\tau_\text{B}}
e^{-iH_\text{XA}\tau_\text{A}} \ket{\downarrow}_\text{X}
\nonumber\\
&={-i}\,\Bigl[
\sA\ketbras{\downarrow}{\uparrow}{\text{A}}\otimes\Bigl(
\ketbras{\uparrow}{\uparrow}{\text{B}}
+\cB\ketbras{\downarrow}{\downarrow}{\text{B}}
\Bigr)
+\sB\,\Bigl(
\ketbras{\downarrow}{\downarrow}{\text{A}}
+\cA\ketbras{\uparrow}{\uparrow}{\text{A}}
\Bigr)\otimes\ketbras{\downarrow}{\uparrow}{\text{B}}
\Bigr].
\end{align}
The sequence of these three processes thereby extracts a state of
the qubits A and B as
\begin{subequations}
\label{eqn:FinalState}
\begin{equation}
\varrho_\text{AB} \to\tilde{\varrho}_\text{AB}
=\mathcal{V}\varrho_\text{AB}\mathcal{V}^\dag/P,
\end{equation}
with the operator
\begin{equation}
\mathcal{V}\equiv R_-L_+R_+.
\end{equation}
\end{subequations}
Note that we retain only those events for which the correct states
are observed on the mediators; other events are discarded. This is
why the states (\ref{eqn:FirstStep}) and (\ref{eqn:FinalState}) are
renormalized, and the normalization constant,
\begin{equation}
P\equiv\Tr_\text{AB}(\mathcal{V}\varrho_\text{AB}\mathcal{V}^\dag),
\end{equation}
is the probability for the entire process depicted in
Fig.~\ref{fig:SeqInts} to be carried out successfully.

The operator $\mathcal{V}$ for this scenario reads
\begin{equation}
\mathcal{V}=2i\sA\cA\sB\,\Bigl(
\cA\sB\ket{\uparrow\downarrow}_\text{AB}
+\sA\ket{\downarrow\uparrow}_\text{AB}
\Bigr)\,\bras{\downarrow\downarrow}{\text{AB}},
\end{equation}
which clearly shows that it extracts the
$\ket{\downarrow\downarrow}_\text{AB}$ component of the initial
state $\varrho_\text{AB}$ and converts it into the pure entangled
state
\begin{equation}
\ket{\Psi}_\text{AB}=\frac{1}{\sqrt{1-\cA^2\cB^2}}\,\Bigl(
\cA\sB\ket{\uparrow\downarrow}_\text{AB}
+\sA\ket{\downarrow\uparrow}_\text{AB} \Bigr),
\label{eqn:GeneratedState}
\end{equation}
i.e.,
\begin{equation}
\varrho_\text{AB}\to\tilde{\varrho}_\text{AB}
=\ketbras{\Psi}{\Psi}{\text{AB}},
\end{equation}
and the yield of this entangled state is given by
\begin{equation}
P=4\sA^2\cA^2\sB^2(1-\cA^2\cB^2)\,
\bras{\downarrow\downarrow}{\text{AB}}\varrho_\text{AB}
\ket{\downarrow\downarrow}_\text{AB}.
\end{equation}

\begin{figure}
\centerline{\includegraphics[width=0.65\textwidth]{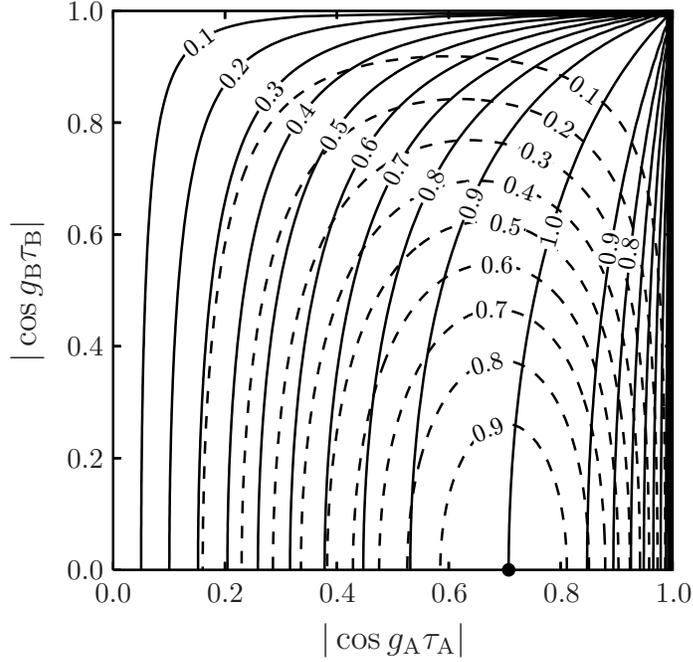}}
\caption{Contour plots of the concurrence $C$ (solid curves) and the yield
$P$ (divided by
$\bras{\downarrow\downarrow}{\text{AB}}\varrho_\text{AB}\ket{\downarrow\downarrow}_\text{AB}$)
(dashed curves) as functions of $g_\text{A}\tau_\text{A}$ and
$g_\text{B}\tau_\text{B}$ for the preparation of the entangled
state $\ket{\Psi}_\text{AB}$ through the processes depicted in
Fig.~\ref{fig:SeqInts}. In this case, $C$ and $P$ are functions
of $|\cos g_\text{A}\tau_\text{A}|$ and $|\cos
g_\text{B}\tau_\text{B}|$. The optimal entanglement preparation with
$C=1$ and $P=\bras{\downarrow\downarrow}{\text{AB}}\varrho_\text{AB}
\ket{\downarrow\downarrow}_\text{AB}$ is realized at the point $(|\cos
g_\text{A}\tau_\text{A}|,|\cos
g_\text{B}\tau_\text{B}|)=(1/\sqrt{2},0)$, indicated by the dot.}
\label{fig:CP}
\end{figure}

The concurrence\cite{ref:Concurrence} of the generated entangled
state $\ket{\Psi}_\text{AB}$ in (\ref{eqn:GeneratedState}) is
readily estimated to be
\begin{equation}
C=\frac{2|\sA\cA\sB|}{\sA^2+\cA^2\sB^2},
\end{equation}
from which we see that $C$ is unity if and only if the parameters
$g_\text{A}\tau_\text{A}$ and $g_\text{B}\tau_\text{B}$ are adjusted
to satisfy the condition
\begin{equation}
\tan^2\!g_\text{A}\tau_\text{A}=\sin^2\!g_\text{B}\tau_\text{B};
\end{equation}
i.e., a \textit{maximally} entangled state can be generated by
tuning the parameters properly. Furthermore, if they are tuned to
satisfy
\begin{equation}
\tan^2\!g_\text{A}\tau_\text{A}=\sin^2\!g_\text{B}\tau_\text{B}=1,
\end{equation}
the maximally entangled state is obtained with yield
$P=\bras{\downarrow\downarrow}{\text{AB}}\varrho_\text{AB}
\ket{\downarrow\downarrow}_\text{AB}$, which is the maximum realizable in the
present scheme.  This means that the entire component of
$\ket{\downarrow\downarrow}_\text{AB}$ contained in the initial
state $\varrho_\text{AB}$ is fully extracted and converted into the
maximally entangled state $\ket{\Psi}_\text{AB}$. Even from the
\textit{maximally mixed} state $\varrho_\text{AB}=\openone_\text{AB}/4$,
the maximally entangled state $\ket{\Psi}_\text{AB}$ can be prepared
with a nonvanishing yield, $P=1/4$. The optimal preparation of the
maximally entangled state between the distant qubits with $C=1$ and
$P=\bras{\downarrow\downarrow}{\text{AB}}\varrho_\text{AB}\ket{\downarrow\downarrow}_\text{AB}$
is thus demonstrated to be possible through only three sequences of
consecutive interactions of the mediator X with the qubits A and
B\@. The parameter dependences of the concurrence $C$ and the yield
$P$ are plotted in Fig.~\ref{fig:CP}.

Similar analyses show that the combinations of the directions (i.e.,
the order of the interactions in each process) other than that
depicted in Fig.~\ref{fig:SeqInts} do not provide the optimal
scheme, and the maximum concurrence and the maximum yield are not
simultaneously realizable. Note again that the parameters
$g_\text{A}\tau_\text{A}$ and $g_\text{B}\tau_\text{B}$ are fixed
throughout all the processes. If it is allowed to vary these
parameters step by step, more efficient schemes can be constructed.
In particular, if the qubits A and B are initialized in the state
$\ket{\uparrow\uparrow}_\text{AB}$ in advance, as described above,
the first two steps in the present scheme are not necessary. The
present scheme is, however, useful in such a situation that it is
not easy to experimentally initialize the system, because it is
still possible to prepare a maximally entangled state directly from
a given mixed state, with a fixed setting. The scheme that is
optimal in any given case depends on the actual experimental setups
and experimental feasibility.

The present scheme makes use of the fact that the interactions given
in (\ref{eqn:Ints}) preserve the excitation number $N$ given in
(\ref{eqn:ExcitationNo}). Such interactions are found in the
literature in the context of quantum information and computation.
For example, generation of entanglement (from a specific pure state)
was demonstrated in a cavity QED experiment \cite{ref:Haroche1997},
in which a cavity mode plays a role similar to that of the mediator
X, and the cavity-atom interaction preserves the excitation number
of the atomic qubits and the cavity mode. For long-range
entanglement, it is quite likely that a photon could act as the
mediator X\@. In connection to this point, the interaction between a
circularly-polarized photon and a $\Lambda$-type atom was recently
proposed in this role\cite{ref:KimbleLambdaAtom2000}. Although this
interaction, between a photon and an atom, is not precisely the
interaction considered here, it also preserves the excitation number
and is a potential system to be used in the experimental
implementation of the present scheme. Applied to such systems, the
present scheme would provide a method of entangling spatially
separated qubits. With the goal of realizing such an experiment, we
should clarify the effects of decoherence of the mediator X, which
may travel a long distance between the qubits A and B, and other
possible disturbances on the qubits during the process.
Investigations of these points are now in progress.

\section*{Acknowledgements}
The authors acknowledge helpful discussions with Professor I. Ohba.
This work is partly supported by a Grant for the 21st Century COE
Program at Waseda University and a Grant-in-Aid for Priority Areas
Research (B) (No.~13135221), both from the Ministry of Education,
Culture, Sports, Science and Technology, Japan, by a Grant-in-Aid
for Scientific Research (C) (No.~14540280) from the Japan Society
for the Promotion of Science, and by the bilateral Italian-Japanese
project 15C1 on ``Quantum Information and Computation'' of the
Italian Ministry for Foreign Affairs.

%\appendix
%\section{First Appendix} %Empty argument \section{} yields `Appendix'.
%
%\section{Second Appendix}


\begin{thebibliography}{99}
%%%%%%%%%%%%%%%%%%%%%%%%%%%%%%%%%%%%%%%%%%%%%%%%%%%%%%%%%%%%%
% Some macros are available for the bibliography:
%  o for general use
%    \JL : general journals                 \andvol : Vol (Year) Page
%  o for individual journal
%    \AJ   : Astrophys. J.           \NC         : Nuovo Cim.
%    \ANN  : Ann. of Phys.           \NPA, \NPB  : Nucl. Phys. [A,B]
%    \CMP  : Commun. Math. Phys.     \PLA, \PLB  : Phys. Lett. [A,B]
%    \IJMP : Int. J. Mod. Phys.      \PRA - \PRE : Phys. Rev. [A-E]
%    \JHEP : J. High Energy Phys.    \PRL        : Phys. Rev. Lett.
%    \JMP  : J. Math. Phys.          \PRP        : Phys. Rep.
%    \JP   : J. of Phys.             \PTP        : Prog. Theor. Phys.
%    \JPSJ : J. Phys. Soc. Jpn.      \PTPS       : Prog. Theor. Phys. Suppl.
% Usage:
%  \PRD{45,1990,345}          ==> Phys.~Rev.\ \textbf{D45} (1990), 345
%  \JL{Nature,418,2002,123}   ==> Nature \textbf{418} (2002), 123
%  \andvol{B123,1995,1020}    ==> \textbf{B123} (1995), 1020
%%%%%%%%%%%%%%%%%%%%%%%%%%%%%%%%%%%%%%%%%%%%%%%%%%%%%%%%%%%%%

\bibitem{ref:QuantInfoCompChuang}
M.~A. Nielsen and I.~L. Chuang,
\textit{Quantum Computation and Quantum Information}
(Cambridge University Press, Cambridge, 2000).

\bibitem{ref:QuantInfoCompZeilinger}
\textit{The Physics of Quantum Information}, ed.\ D.
Bouwmeester, A. Ekert and A. Zeilinger
(Springer-Verlag, Heidelberg, 2000).\\
C.~H. Bennett and D.~P. DiVincenzo, \JL{Nature (London),404,2000,247}.\\
A. Galindo and M.~A. Martin-Delgado, \JL{Rev. Mod. Phys.,74,2002,347}.

\bibitem{ref:Peres}
A. Peres, \textit{Quantum Theory: Concepts and Methods}
(Kluwer Academic Publishers, Dordrecht, 1995).

\bibitem{ref:Haroche1997}
E. Hagley, X. Ma\^itre, G. Nogues, C. Wunderlich, M. Brune,
J.~M. Raimond and S. Haroche, \PRL{79,1997,1}.\\
J.~M. Raimond, M. Brune and S. Haroche, \JL{Rev. Mod. Phys.,73,2001,565}.

\bibitem{ref:Bergou1997}
J.~A. Bergou and M. Hillery, \PRA{55,1997,4585}.\\
A. Messina, \JL{Eur. Phys. J. D,18,2002,379}.\\
D.~E. Browne and M.~B. Plenio, \PRA{67,2003,012325}.

\bibitem{ref:CiracZollerEntGeneDistant}
C. Cabrillo, J.~I. Cirac, P. Garc\'ia-Fern\'andez and P. Zoller,
\PRA{59,1999,1025}.\\
L.-M. Duan, M.~D. Lukin, J.~I. Cirac and P. Zoller,
\JL{Nature (London),414,2001,413}.\\
D.~E. Browne, M.~B. Plenio and S.~F. Huelga, \PRL{91,2003,067901}.

\bibitem{ref:EntanglingDistantAtoms2003}
X.-L. Feng, Z.-M. Zhang, X.-D. Li, S.-Q. Gong and Z.-Z. Xu,
\PRL{90,2003,217902}.\\
L.-M. Duan and H.~J. Kimble, \PRL{90,2003,253601}.

\bibitem{ref:ZeilingerTeleportationDanube}
R. Ursin, T. Jennewein, M. Aspelmeyer, R. Kaltenbaek, M. Lindenthal,
P. Walther and A. Zeilinger, \JL{Nature (London),430,2004,849}.

\bibitem{ref:PurificationBennettPRL1996}
C.~H. Bennett, G. Brassard, S. Popescu, B. Schumacher, J.~A. Smolin
and W.~K. Wootters, \PRL{76,1996,722} [Errata; \textbf{78} (1997), 2031].\\
C.~H. Bennett, D.~P. DiVincenzo, J.~A. Smolin and W.~K. Wootters,
\PRA{54,1996,3824}.

\bibitem{ref:TYamamotoNature}
T. Yamamoto, M. Koashi, \c{S}.~K. {\"O}zdemir and N. Imoto,
\JL{Nature (London),421,2003,343}.\\
Z. Zhao, T. Yang, Y.-A. Chen, A.-N. Zhang and J.-W. Pan,
\PRL{90,2003,207901}.\\
A. Vaziri, J.-W. Pan, T. Jennewein, G. Weihs and A. Zeilinger,
\PRL{91,2003,227902}.

\bibitem{ref:qpf}
H. Nakazato, T. Takazawa and K. Yuasa, \PRL{90,2003,060401}.

\bibitem{ref:qpfe}
H. Nakazato, M. Unoki and K. Yuasa, \PRA{70,2004,012303}.

\bibitem{ref:qpfeLidar}
L.-A. Wu, D.~A. Lidar, and S. Schneider, \PRA{70,2004,032322}.

\bibitem{ref:QZE}
See, for example,
B. Misra and E.~C.~G. Sudarshan, \JMP{18,1977,756}.\\
H. Nakazato, M. Namiki and S. Pascazio,
\JL{Int. J. Mod. Phys. B,10,1996,247}.\\
D. Home and M.~A.~B. Whitaker,
\JL{Ann. of Phys.,258,1997,237}.\\
P. Facchi and S. Pascazio, in \textit{Progress in Optics},
ed.\ E. Wolf (Elsevier, Amsterdam, 2001), Vol.\ 42, p.\ 147.

\bibitem{ref:qpfes}
G. Compagno, A. Messina, H. Nakazato, A. Napoli, M. Unoki and
K. Yuasa, \PRA{70,2004,052316}.

\bibitem{ref:Concurrence}
W.~K. Wootters, \PRL{80,1998,2245}.

\bibitem{ref:KimbleLambdaAtom2000}
W. Lange and H.~J. Kimble, \PRA{61,2000,063817}.\\
J. Hong and H.-W. Lee, \PRL{89,2002,237901}.\\
B. Sun, M.~S. Chapman and L. You, \PRA{69,2004,042316}.
\end{thebibliography}
\end{document}